\documentclass[aps,prl,twocolumn,groupedaddress,showpacs,amsmath,amssymb]{revtex4}
\usepackage{epsfig}
\usepackage{amssymb}
\usepackage{amsmath}
\usepackage{graphicx}
\usepackage{amsfonts}
\usepackage{epsfig}
\usepackage{revsymb}
\usepackage{bm}
\usepackage{amsmath}
\usepackage{amssymb}
\usepackage{latexsym}
\usepackage{thumbpdf}

\begin{document}

\title{Dissipative dynamics of the Josephson effect in the binary
Bose-condensed mixtures}

\author{S. N. Burmistrov}

\affiliation{Kurchatov Institute, 123182 Moscow, Russia}


\begin{abstract}
\par
The dissipative dynamics of a pointlike Josephson junction in binary
Bose-condensed mixtures is analyzed within the framework of the model of a
tunneling Hamiltonian. The transmission of unlike particles across a junction is
described by the different transmission amplitudes. The effective action that
describes the dynamics of the phase differences across the junction for each of
two condensed components is derived employing the functional integration method.
In the low-frequency limit the dynamics of a Josephson junction can be described
by two coupled equations in terms of the potential energy and dissipative Rayleigh
function using a mechanical analogy. The interplay between mass currents of each
mixture component appears in the second-order term in the tunneling amplitudes due
to interspecies hybridizing interaction. The asymmetric case of the binary
mixtures with the different concentration and order parameters is considered as
well.
\end{abstract}

\pacs{03.75.Lm, 67.85.Fg, 74.50+r}

\maketitle

\section{INTRODUCTION}
\par
Recently, experimental study of multicomponent Bose-Einstein condensates has made
a substantial progress. The study of multiple atomic condensates is intriguing
since they can produce a laboratory mixture of distinguishable boson superfluids
at sufficiently low temperatures. A considerable amount of theoretical work has
been devoted to binary Bose-condensed mixtures, focusing, mainly, on the
mean-field description of trapped binary mixtures \cite{Es}, stability and phase
separation \cite{Lar,Ti,Ao}, collective excitations \cite{Bus,Gra,Pu}, condensate
depletions \cite{Eck},  and quantum merging of two different condensates
\cite{Zap,Yi,Meb}.
\par
In this connection we will consider here the dissipative and interference aspects
of the Josephson effect in the binary Bose-condensed mixtures. The Josephson
effect, first predicted and discovered for two superconductors separated with a
thin insulator layer, is a macroscopic quantum phenomenon in a condensed medium.
The dynamics of the effect is described in terms of the difference between the
phases of the superconductors, playing the role of a macroscopic quantum variable.
In spite of wide application of the effect in devices for extremely
high-sensitivity measurements of currents, voltages, and magnetic fields the
Josephson effect is still of interest in the fundamental modern physics. Like
superconductors and Fermi superfluids, the Josephson effect is also inherent in
Bose superfluids \cite{Su,Ho,Na}. The effect has been observed  by the mass flow
of superfluid $^4$He through nanoscale apertures coupling two bulk superfluid
reservoirs. The Josephson tunneling junction in ultracold dilute atomic gases is
formed with a laser separating two Bose-Einstein condensates \cite{Al,Le}.
\par
In general, the Josephson effect can include both the so-called internal effect
for the atoms in different hyperfine states much as NMR phenomena in $^3$He
\cite{Leg} and the conventional case of two Bose condensates separated with a
potential barrier which acts as a tunneling junction. A lot of work
\cite{Mi,Sm,Za,St,Ra,Vi,Sme,Chi,Wi,Mei,Ba,Xi,Tru,Ni,Bon} has been done in the
latter case as a direct analogy with conventional superconductors. Those studies
dealt with the one-component Bose-Einstein condensates alone.
\par
Multicomponent Bose-Einstein condensates are also a very interesting subject for
studying various macroscopic tunneling phenomena \cite{Xu,Sa,Jul,Ma,Nad}. One may
expect novel and richer manifestations of the Josephson effect. For the system of
two Bose-condensed mixtures connected with a weakly coupled junction, the dynamics
of the Josephson effect should be governed by the difference between the phases
for each Bose-condensed component of a mixture. In other words, two relative
phases $\varphi _1$ and $\varphi _2$ must be involved into consideration. In
addition, we must take into account the different tunneling transition amplitudes
$I_1$ and $I_ 2$ across the junction for various bosonic atoms of masses $m_1$ and
$m_2$ composing the mixture. This results in two Josephson currents associated
with the mass flow of each component of a mixture across the junction. From the
general point of view one may expect interference and coupling between the
Josephson currents. The dynamics of the Josephson effect in the vicinity of the
phase separation of a mixture is an additional motivation for studying  binary
condensates. Furthermore, the dissipative aspects of the Josephson dynamics in
binary condensed mixtures have not yet received a proper and wide investigation.
\par
The dissipative effects and dephasing of the Josephson oscillations come from the
coupling between the macroscopic relative phase variable and the infinite number
of the microscopic degrees of freedom. The successive method of eliminating
microscopic degrees of freedom from the Hamiltonian was developed first for the
superconducting Josephson systems \cite{Am,La,Ec}. Later that functional
integration approach \cite{Mei,Ba} and the Keldysh Green function method
\cite{Tru} were extended and applied to studying dissipative and nonequilibrium
Josephson dynamics in the one-component Bose-condensed systems.
\par
In this paper, we will generalize the energy dissipation effects in the Josephson
dynamics to the case of the binary Bose-condensed systems, employing functional
integration approach of Ref.~\cite{Ba}. We will derive the expression for the
effective action depending on two relative phases $\varphi _1$ and $\varphi _2$
between two condensed mixtures connected by a pointlike tunneling junction. The
response functions in the effective action give the full information on the
dynamics of the junction. The low-frequency expansion of the response functions
allows us to determine two coupled Josephson equations for the relative phases
$\varphi _1$ and $\varphi _2$, Josephson energy $U(\varphi _1,\,\varphi _2)$ and
dissipative Rayleigh function $R(\dot{\varphi} _1,\,\dot{\varphi} _2)$. Of course,
we consider the region of the parameters in which the homogeneous state of the
both mixtures in the left-hand and right-hand bulks is stable and the mixtures are
not phase-separated.

\section{EFFECTIVE ACTION}
\par
We keep in mind the case of a pointlike and weakly coupled junction between two
macroscopic infinite reservoirs containing binary condensed mixtures. In addition,
we neglect the feedback effect of the junction on the mixtures and assume that
both the mixtures are always in the thermal equilibrium state. The image of the
system is two bulks with one common point through which the transmission of
particles is possible with the different tunneling amplitudes $I_1$ and $I_2$
depending on the type of particles.
\par
So, our starting point is the so-called tunneling Hamiltonian ($\hbar =1$, volume $V=1$)
\begin{equation}\label{f01}
H=H_{l}+H_{r}+H_{U}+H_{t}\, ,
\end{equation}
where $H_{l, \, r}$ describes the bulk binary Bose-condensed mixture on the left-hand and right-hand sides,
respectively,
\begin{gather*}
H_{l}= \sum\limits_{i =1,\, 2}\int\!\! d^{3}r\,\Psi _{i ,\, l}^{\dagger}\!\left(
-\frac{\nabla ^2}{2m_{i}}-\mu _{i} \right)\!\Psi _{i ,\, l}
\\
+\frac{1}{2}\sum\limits_{i,\, k =1,\, 2}u_{ik ,\, l }\int\!\! d^{3}r\,\Psi _{i ,\,
l}^{\dagger}\Psi _{k ,\, l}^{\dagger}\Psi _{k ,\, l}\Psi _{i ,\, l} .\nonumber
\end{gather*}
Here $i$ and $k$ take 1 or 2 and denote one of the components of a mixture
composed with particles of mass $m_1$ and $m_2$. The coupling between particles is
specified by the constants $u_{ik,\, l }=u_{ki,\, l }$ which can be expressed by
means of the $s$-scattering length $a_{ik,\, l}$ according to $u_{ik,\, l}=2\pi
a_{ik,\, l}(m_{i}^{-1}+m_{k}^{-1})$. The same expressions with the substitution
$l\rightarrow r$ refer to the mixture on the right-hand side.
\par
The energy, associated with varying the number of particles on the left-hand and right-hand sides,
\begin{equation*}
H_U=\frac{1}{2}\sum\limits_{i,\, k =1,\, 2}\frac{N_{i,\, l}-N_{i,\, r}}{2}\,
U_{ik}\frac{N_{k,\, l}-N_{k,\, r}}{2}\, ,
\end{equation*}
is analogous to the capacity energy of a junction in the case of superconductors.
The constants $U_{ik}$ can be connected with the second derivatives of the total
energy
\begin{gather*}
E=E[(N_{1,\, l},\, N_{2,\, l}),\, (N_{1,\, r},\, N_{2,\, r})]
\\
=E_l(N_{1,\, l},\, N_{2,\, l})+E_r(N_{1,\, r},\, N_{2,\, r})
\end{gather*}
with respect to the relative change in the number of the particles across the
junction,
\begin{gather*}
U_{ik}= \frac{\partial ^2E}{\partial N_{i, \, l}\partial N_{k, \,
l}}+\frac{\partial ^2E}{\partial N_{i, \, r}\partial N_{k, \, r}}
\\
=\frac{\partial ^2E_l}{\partial N_{i, \, l}\partial N_{k, \, l}}+\frac{\partial
^2E_r}{\partial N_{i, \, r}\partial N_{k, \, r}}\, ,
\end{gather*}
with the obvious symmetrical  relation $U_{ik}=U_{ki}$. The constants $U_{ik}$ can
usually be estimated also as
\begin{equation*}
U_{ik}=\frac{\partial\mu _{i,\, l}}{\partial N_{k, \, l}}+\frac{\partial\mu _{i,\,
r}}{\partial N_{k, \, r}}\, .
\end{equation*}
\par
The term $H_U$ describes the point that the energy of the system on the whole may
depend on the relative numbers of particles from the left-hand and right-hand
bulks. In order to avoid an instability of the total system against an infinite
growth of the number of particles on the left-hand or right-hand sides, it is
necessary to suppose that $H_U>0$ for any variations $N_{1,\, l}-N_{1,\, r}$ and
$N_{2,\, l}-N_{2,\, r}$. In other words, the matrix of coefficients $U_{ik}$
should be positively determined, i.e., $\|U_{ik}\|>0$. The total number of the
particles of the type labelled by $i$ in each bulk $l$ or $r$ is given by
\begin{equation*}
N_{i,\, l}=\int\!\! d^{3}r\,\Psi _{i,\, l}^{\dagger}\Psi _{i,\,
l}\,\,\,\text{and}\,\, (l\rightarrow r)\, .
\end{equation*}
\par
The last term,
\begin{gather*}
H_t=-\!\!\int _{\bm{r}\in l,\, \bm{r}'\in r}\!\!\! d^3r\, d^3r'\,\bigl[\Psi _{1,\,
l}^{\dagger}(\bm{r})I_1(\bm{r},\,\bm{r}')\Psi _{1,\, r}(\bm{r}')
\\
+\Psi _{2,\, l}^{\dagger}(\bm{r})I_2(\bm{r},\,\bm{r}')\Psi _{2,\, r}(\bm{r}')+ \text{H.c.}\bigr],
\end{gather*}
is responsible for the transitions of particles from the right-hand to the
left-hand bulk and \textit{vice versa}. In general, the transition amplitudes
$I_1$ and $I_2$ are different for the various species of the particles composing
the mixture. We consider here the simplest case of a pointlike junction, i.e.,
 $$
I_{i}(\bm{r},\,\bm{r}')=I_{i}\delta (\bm{r})\delta (\bm{r}'), \;\;\; i =1,\, 2.
 $$
\par
To study the properties of the system described by Eq.~(\ref{f01}), we calculate
the partition function $Z$ using the analogy of the superconducting junction
\cite{Am,La,Ec} and the approaches employed for the Bose junction \cite{Mei,Ba},
\begin{equation*}
Z=\int\mathcal{D}^2\Psi _{1,\, l}\mathcal{D}^2\Psi _{2,\, l}\mathcal{D}^2\Psi
_{1,\, r}\mathcal{D}^2\Psi _{2,\, r}\exp [-S_E],
\end{equation*}
where the action $S_E$, defined on the imaginary (Matsubara) time $\tau$, reads
\begin{gather*}
S_E=\int _{-\beta /2}^{\beta /2}d\tau\, L_E ,
\\
L_E = \int d^3r\!\!\sum\limits_{i,\, k =1,\, 2}\left(\Psi ^{\dagger}_{i,\,
l}\frac{\partial}{\partial\tau}\Psi _{i,\, l}+\Psi ^{\dagger}_{i,\,
r}\frac{\partial}{\partial\tau}\Psi _{i,\, r}\right) +H
\end{gather*}
and, as usual, $\beta=1/T$ is an inverse temperature.
\par
To eliminate the quartic terms in the action, which come from the energy $H_U$, we
employ the Hubbard-Stratonovich procedure by introducing additional gauge fields
$V_1(\tau )$ and $V_2(\tau )$ in analogy with the so-called plasmon gauge field in
metals,
\begin{gather*}
\exp\Big(-\!\int d\tau\, H_U\Big) =\int\!\mathcal{D}V_1(\tau
)\,\mathcal{D}V_2(\tau )
\\
\exp\Big[-\!\!\int d\tau\Big(\sum\limits _{i,k=1,2}\frac{
V_{i}\widehat{U}_{ik}^{-1}V_{k}}{2}+i\sum\limits _{i=1,2}\frac{N_{i,\, l}-N_{i,\,
r}}{2}\, V_{i}\Big)\Big] ;
\\
\int\!\mathcal{D}V_1(\tau )\,\mathcal{D}V_2(\tau )\exp\Big[-\int d\tau\sum\limits
_{i,k=1,2}\frac{V_i\widehat{U}_{ik}^{-1}V_k}{2}\Big]=1 .
\end{gather*}
Here  $\widehat{U}_{ik}^{-1}$ is an inverse matrix for the $2\times 2$ matrix
$U_{ik}$
\begin{eqnarray*}
\widehat{U}_{ik}^{-1}=\frac{1}{U_{11}U_{22}-U_{12}U_{21}}\left(
\begin{array}{cc}
U_{22} & -U_{12}
\\
-U_{21} & U_{11}
\end{array}
\right) .
\end{eqnarray*}
\par
After the introduction of the fields $V_1(\tau)$ and $V_2(\tau)$ the partition
function $Z$ takes the form
 $$
Z=\int\!\mathcal{D}V_1\,\mathcal{D}V_2\prod _{i=1,\, 2}\mathcal{D}^2\Psi
_{i,\,l}\mathcal{D}^2\Psi _{i,\, r}\exp [-\tilde{S}_E],
 $$
where
 $$
\tilde{S}_E=\tilde{S}_l+\tilde{S}_r+\int d\tau\, H_t +\frac{1}{2}\sum\limits
_{i,k=1,2}\int d\tau\, V_{i}\widehat{U}_{ik}^{-1}V_{k} .
 $$
Here $\tilde{S}_l$ and $\tilde{S}_r$ denote
\begin{gather*}
\tilde{S}_l=S_l+\frac{i}{2}\sum\limits_{i=1,\, 2}\int d\tau\, d^3r\,\Psi _{i,\,
l}^{\dagger}V_{i}(\tau ) \Psi _{i,\, l} ,
\\
\tilde{S}_r=S_r-\frac{i}{2}\sum\limits_{i=1,\, 2}\int d\tau\, d^3r\,\Psi _{i,\,
r}^{\dagger}V_{i}(\tau ) \Psi _{i,\, r} .
\end{gather*}
In essence, this replacement looks like the renormalization of the chemical
potentials for each component of a mixture on the left and right-hand sides of the
junction,
\begin{gather*}
\mu _{i,\, l}\rightarrow \mu _{i,\, l}-iV_i (\tau)/2\, ,
\\
\mu _{i,\, r}\rightarrow \mu _{i,\, r}+iV_i(\tau)/2\, .\nonumber
\end{gather*}
\par
At this point it is advantageous to perform a gauge transformation of the field
operators $\Psi ^{\dagger}$ and $\Psi$, which makes the future Green functions
real. This is achieved by introducing the phases $\varphi _{i,\, l}$ and $\varphi
_{i,\, r}$ according to
\begin{gather*}
\Psi _{i,\, l\, (i,\, r)}\rightarrow \exp\bigl[i\varphi _{i,\, l\, (i,\,
r)}(\tau)\bigr]\Psi _{i,\, l\, (i,\, r)} ,
\\
\Psi ^{\dagger}_{i,\, l\, (i,\, r)}\rightarrow \exp\bigl[-i\varphi _{i,\, l\,
(i,\, r)}(\tau)\bigr]\Psi ^{\dagger}_{i,\, l\, (i,\, r)} ,
\end{gather*}
and by imposing the conditions $\dot{\varphi} _{i,\, l}=-V_{i}(\tau )/2$ and
$\dot{\varphi} _{i,\, r}=V_{i}(\tau )/2$. Thus, we arrive at the first Josephson
relations for the relative phases $\varphi _1(\tau )$ and $\varphi _2(\tau )$:
\begin{equation}\label{f16}
\dot{\varphi} _{i}(\tau )=V_{i}(\tau ) \;\;\;\text{and} \;\;\; \varphi
_{i}=\varphi _{i,\, r }-\varphi _{i,\, l}\;\;\; (i=1,\, 2).
\end{equation}
That we have achieved is only a formal elimination of the explicit dependence of
the chemical potentials $\mu _{i,\, l}$ and $\mu _{i,\, r}$ upon the time $\tau$.
On the other hand, the tunneling amplitudes $I_1$ and $I_2$ acquire additional
factors depending on the phase differences $\varphi _1(\tau )$ and $\varphi
_2(\tau )$ across the junction,
\begin{equation*}
I_1\rightarrow \tilde{I}_1=I_1\text{e}^{i\varphi _1(\tau)}\,\,\, \text{and}\;\;\;
I_2\rightarrow \tilde{I}_2=I_2\text{e}^{i\varphi _2(\tau)} .
\end{equation*}
Hence we arrive at
\begin{gather*}
Z=\int\!\mathcal{D}V_1\,\mathcal{D}V_2\prod _{i=1,\, 2}\mathcal{D}^2\Psi _{i,\,
l}\mathcal{D}^2\Psi _{i,\, r}\, e^{-S},
\\
S=S_0+\!\int d\tau\,\tilde{H}_t +\int d\tau\sum\limits_{i,\, k =1,\, 2}
V_{i}\frac{\widehat{U}_{ik}^{-1}}{2}V_{k},
\end{gather*}
where $S_0=S_l+S_r$. The tunneling term $\tilde{H}_t$ is given by
 $$
\tilde{H}_t=- \!\!\!\!\int\limits_{\bm{r}\in l,\, \bm{r}'\in r}\!\!\! d^3r\,
d^3r'\sum\limits _{i=1,\, 2} \bigl[\Psi _{i,\,
l}^{\dagger}\tilde{I}_{i}(\bm{r},\bm{r}';\,\tau)\Psi _{i,\, r}+ \text{H.c.}\bigr]
.
 $$
\par
Next, one must integrate over fields $\Psi ^{\dagger}$ and $\Psi $ in order to
obtain the effective action $S_{\text{eff}}$ depending on $V_{1}(\tau )$ and
$V_{2}(\tau )$ alone. In calculations we treat $\tilde{H}_t$ as a perturbation and
restrict ourselves by second-order perturbation in the tunneling amplitudes $I_1$
and $I_2$. Omitting the term independent of $V_{i}(\tau )$ and employing the
Josephson relations $V_{i}(\tau )=\dot{\varphi}_i(\tau )$, we find the effective
action as
 $$
S_{\text{eff}}[\varphi _1,\varphi _2]= \int\! d\tau\!\Bigl[\sum\limits_{i,k=1,2}\!
\dot{\varphi}_{i}\frac{\widehat{U}_{ik}^{-1}}{2}\dot{\varphi}_{k}+\langle\tilde{H}_t\rangle
_0 -\frac{\langle\langle\tilde{H}_t^2\rangle\rangle _0 }{2}\Bigr] .
 $$
Here $\langle A\rangle _0$ means the averaging over decoupled action $S_0=S_l+S_r$
corresponding to $H_0=H_l+H_r$, i.e.,
 $$
\langle\tilde{H}_t\rangle =\frac{\langle\tilde{H}_t\, e^{-S_0}\rangle}{\langle
e^{-S_0}\rangle}\;\;\text{and}\;\;\langle\langle\tilde{H}_t^2\rangle\rangle _0
=\langle\tilde{H}_t^2\rangle _0 -\langle\tilde{H}_t\rangle _0^2\, .
 $$
It is obvious that
\begin{gather*}
\langle\tilde{H}_t\rangle _0=\langle\tilde{H}_{1t}\rangle
_0+\langle\tilde{H}_{2t}\rangle _0\,
\\
\langle\langle\tilde{H}_t^2\rangle\rangle _0
=\langle\langle\tilde{H}_{1t}^2\rangle\rangle
_0+2\langle\langle\tilde{H}_{1t}\tilde{H}_{2t}\rangle\rangle _0
+\langle\langle\tilde{H}_{2t}^2\rangle\rangle _0\, .
\end{gather*}
The first-order terms in the tunneling transparency are obviously decoupled. Since
$\tilde{H}_{1t}\tilde{H}_{2t}\sim I_1I_2$, second-order terms will result in the
coupling and interference between the mass currents of atom species 1 and 2.
Averaging over the $l$ and $r$ variables is independent of each other.
\par
In the course of calculation we follow the Bogoliubov method of separating the
field operators into the condensate $C$ and noncondensate $\Phi$ fractions, i.e.,
 $$
\Psi _{i,\, l(i,\, r)}=C_{i,\, l(i,\, r)}+\Phi _{i,\, l(i,\, r)},
 $$
with the conventional relation $C_{i,\, l(i,\, r)}=\sqrt{\, n_{i,\, l(i,\, r)}}$
where ${n_{i,\, l(i,\, r)}}$ is the density of particles labelled with $i=1$, $2$
in the condensate fraction in the left-hand and right-hand bulks, respectively.
\par
In a binary Bose-condensed mixture the Green function $\widehat{G}(\omega
_n,\bm{p})$ represents a block $4\times 4$ matrix. The Green function can readily
be found from the inverse matrix whose Fourier representation in the approximation
of a weakly interacting two-component Bose-condensed gas mixture is given by
\begin{widetext}
\begin{equation*}
\widehat{G}^{-1}(\omega _n,\bm{p})\!\!=\!\!\left(
\begin{array}{cc}
\!\!\widehat{G}_{11}^{-1} & \hat{\Delta}_{12}\!\!
\\
\!\!\hat{\Delta}_{12} & \widehat{G}_{22}^{-1}\!\!
\end{array}
\right)\!\!=\!\!\left(
\begin{array}{cccc}
\!\! -i\omega _n+\eta _1 +\Delta _{11}\!\! & \Delta _{11} & \Delta _{12} & \Delta
_{12}
\\
\Delta _{11} & \!\! i\omega _n+\eta _1 +\Delta _{11}\!\! & \Delta _{12} & \Delta
_{12}
\\
\Delta _{12} & \Delta _{12} & \!\! -i\omega _n+\eta _2 +\Delta _{22}\!\! & \Delta
_{22}
\\
\Delta _{12} & \Delta _{12} & \Delta _{22} & \!\! i\omega _n+\eta _2 +\Delta
_{22}\!\!
\end{array}
\right)\! .
\end{equation*}
\end{widetext}
Here $\omega _n=2\pi nT$ is the Matsubara frequency  and $\bm{p}$ is the momentum.
Also we have introduced the following notations for the free-particle energies
\begin{equation*}
\eta _1=\eta _1(\bm{p})=\bm{p}^2/2m_1\, ,\;\;\;\eta_2=\eta
_2(\bm{p})=\bm{p}^2/2m_2\, ;
\\
\end{equation*}
and for the order parameters
\begin{equation*}
\Delta _{11}=u_{11}n_{1}\, ,\;\; \Delta _{22}=u_{22}n_{2}\,
,\;\;\Delta _{12}=u_{12}\sqrt{n_{1}n_{2}}\, .
\end{equation*}
Accordingly, for the direct matrix Green function
\begin{equation*}
\widehat{G}(\omega _n,\,\bm{p})=\left(
\begin{array}{cc}
\widehat{G}_{11} & \widehat{G}_{12}
\\
\widehat{G}_{21} & \widehat{G}_{22}
\end{array}
\right) =\left(
\begin{array}{cccc}
G_{11} & F _{11} & G_{12} & F_{12}
\\
F_{11}^{\dagger} & \overline{G}_{11} & F_{12}^{\dagger} & \overline{G}_{12}
\\
G_{21} & F _{21} & G_{22} & F_{22}
\\
F_{21}^{\dagger} & \overline{G}_{21} & F_{22}^{\dagger} & \overline{G}_{22}
\end{array}
\right) ,
\end{equation*}
we arrive at the following components of the matrix:
\begin{widetext}
\begin{gather*}
G_{11}(\omega _n,\,\bm{p})=\overline{G}_{11}(-\omega _n,\,\bm{p})=\frac{(i\omega
_n +\eta _1+\Delta _{11})(\omega _n^2+\varepsilon _2^2)-2\Delta _{12}^2\eta
_2}{(\omega _n^2+\omega _1^2)(\omega _n^2+\omega _2^2)}
\\
F_{11}(\omega _n,\,\bm{p})=F_{11}^{\dagger}(-\omega _n,\,\bm{p})=\frac{-\Delta
_{11}(\omega _n^2+\varepsilon _2^2)+2\Delta _{12}^2\eta _2}{(\omega _n^2+\omega
_1^2)(\omega _n^2+\omega _2^2)}
\\
G_{12}(\omega _n,\,\bm{p})=\overline{G}_{12}(-\omega _n,\,\bm{p})=-\,\frac{\Delta
_{12}(i\omega _n+\eta _1)(i\omega _n+\eta _2)}{(\omega _n^2+\omega _1^2)(\omega _n
^2+\omega _2^2)}
\\
F_{12}(\omega _n,\,\bm{p})=F_{12}^{\dagger}(-\omega _n,\,\bm{p})=-\,\frac{\Delta
_{12}(i\omega _n+\eta _1)(-i\omega _n+\eta _2)}{(\omega _n^2+\omega _1^2)(\omega
_n ^2+\omega _2^2)}
\end{gather*}
\end{widetext}
The lower part of the Green function matrix is determined by permutation
$1\leftrightarrows 2$ and $\Delta _{21}=\Delta _{12}$. We have introduced above
the similar abbreviations for either of two mixtures, using the Bogoliubov
nomenclature
 $$
\varepsilon _1^2=\eta _1^2+2\Delta _{11}\eta _1\;\;\;\text{and}\;\;\;\varepsilon
_2^2=\eta _2^2+2\Delta _{22}\eta _2
 $$
where $\eta _{1,\, 2}=\bm{p}^2/2m_{1,\, 2}$ is the free-particle energy. The
energies $\varepsilon _1$ and $\varepsilon _2$ are the Bogoliubov energies of each
component of a mixture taken separately. In the mixture the interspecies
interaction $u_{12}$ hybridizes these two modes, resulting in two familiar
branches of elementary excitation spectrum, e.g., \cite{Ti,Eck}
\begin{equation*}
\omega _{1, 2}^2(p)=\frac{1}{2}\Bigl(\varepsilon _1^2+\varepsilon
_2^2\pm\sqrt{(\varepsilon _1^2-\varepsilon _2^2)^2+16\Delta _{12}^2\eta _1\eta
_2}\,\,\Bigr)
\end{equation*}
with the crossover to the sound-like dispersion $\omega _{1, 2}=p\, c_{1, 2}$ at
small $ p\rightarrow 0$ momentum. The sound velocities $c_{1,2}$ in a mixture are
determined by the well-known relations as well \cite{Ti,Eck}
 $$
c _{1, 2}^2=\frac{1}{2}\Bigl[\frac{\Delta _{11}}{m_1}+\frac{\Delta
_{22}}{m_2}\pm\sqrt{\Bigl(\frac{\Delta _{11}}{m_1}-\frac{\Delta
_{22}}{m_2}\Bigr)^2+4\frac{\Delta _{12}^2}{m _1m _2}}\,\,\Bigr] .
 $$
The inequalities $u_{11}u_{22}>u_{12}^2$ and $c_{1,2}^2>0$ are certainly supposed
to guarantee the stability of a mixture against its demixing.
\par
Taking into account (\ref{f16}), we arrive finally at the following generalization
of the effective action compared with  that in the one-component condensed system
\cite{Ba}
\begin{gather}
S_{\text{eff}}[\varphi _1(\tau ), \varphi _2(\tau )]=\int
d\tau\,\Bigl[\sum\limits_{i,k=1,2}\dot{\varphi}
_i(\tau)\frac{\widehat{U}_{ik}^{-1}}{2}\dot{\varphi}_k(\tau)\nonumber
\\
-2I_1\sqrt{n_{1\, l}n_{1\, r}}\,\cos\varphi _1(\tau) -2I_2\sqrt{n_{2\, l}n_{2\,
r}}\,\cos\varphi _2(\tau)\Bigr] \nonumber
\\
-\!\!\sum _{i,k=1,2}\!\! I_iI_k\int d\tau\, d\tau '\,\Bigl[\alpha _{ik}(\tau -\tau
')\cos\Bigl(\varphi _i(\tau)-\varphi _k(\tau ')\Bigr) \nonumber
\\
+\beta _{ik}(\tau -\tau ')\cos\Bigl(\varphi _i(\tau)+\varphi _k(\tau
')\Bigr)\Bigr] . \label{f18}
\end{gather}
Here $\alpha _{ik}$ and $\beta _{ik}$ are the so-called response functions which
can be written using the Green functions:
\begin{gather*}
\alpha _{ik}(\tau)=\sqrt{n_{i\, l}n_{k\, l}}\, g_{ik,\, r}^+(\tau ) + \sqrt{n_{i\,
r}n_{k\, r}}\, g_{ik,\, l}^+(\tau ) +\mathcal{G}_{ik}(\tau ) ,
\\
\beta _{ik}(\tau)=\sqrt{n_{i\, l}n_{k\, l}}\, f_{ik,\, r}^+(\tau ) + \sqrt{n_{i\,
r}n_{k\, r}}\, f_{ik,\, l}^+(\tau ) +\mathcal{F}_{ik}(\tau ).
\end{gather*}
The Green functions are calculated at the junction point, i.e., at $\bm{r}=0$ and
$\bm{r}'=0$,
\begin{gather*}
g_{ik,\, l(r)}^+(\tau )=\frac{1}{2}[g_{ik,\, l(r)}(\tau )+\bar{g}_{ik,\,
l(r)}(\tau )]
\\
=\int\frac{d^3p}{2(2\pi )^3}\,\bigl[G_{ik,\, l(r)}(\bm{p},\, \tau
)+\overline{G}_{ik,\, l(r)}(\bm{p},\, \tau )\bigr] ,
\\
f_{ik,\, l(r)}^+(\tau )=\int\frac{d^3p}{2(2\pi )^3}\,\bigl[ F_{ik,\,
l(r)}(\bm{p},\, \tau )+F_{ik,\, l(r)}^{\dagger}(\bm{p},\, \tau )\bigr] ;
\\
\mathcal{G}_{ik}(\tau )=g_{ik,\, l}(\tau )\bar{g}_{ik,\, r}(\tau )+\bar{g}_{ik,\,
l}(\tau ) g_{ik,\, r}(\tau ) ,
\\
\mathcal{F}_{ik}(\tau )=f_{ik,\, l}(\tau )f_{ik,\, r}^{\dagger}(\tau )+f_{ik,\,
l}^{\dagger}(\tau ) f_{ik,\, r}(\tau ) .
\end{gather*}
\par
In order to comprehend the dynamics of the relative phase differences $\varphi _1$
and $\varphi _2$ across the junction, we should analyze the behavior of the
response functions $\alpha _{ik}(\tau )$ and $\beta _{ik}(\tau)$ as a function of
time. Note that the contribution of the terms $\mathcal{G}_{ik}$ and
$\mathcal{F}_{ik}$ to the response functions $\alpha _{ik}$ and $\beta _{ik}$ is
much smaller than that of the first two others \cite{Ba}. The order-of-magnitude
smallness is about a ratio of the noncondensate density to the condensate density
or about gas parameter $(na^3)^{1/2}\ll 1$. Below, analyzing $\alpha _{ik}$ and
$\beta _{ik}$, we will concentrate our attention on the first two terms which can
be attributed to the condensate-noncondensate tunneling processes.

\section{THE RESPONSE FUNCTIONS}
\par
The calculation of the response functions in the general form in a mixture is a
complicated problem. Keeping in mind the study of the low-frequency dynamics of a
junction, we will restrict our calculation by analyzing the behavior of the
response functions on the long-time scale. This means that we should find the
low-frequency decomposition of the response functions in the Matsubara frequencies
$\omega _n$. In fact, we imply the inequality $|\omega _n|\ll\omega _1$, $\omega
_2$. Next, we will employ the procedure of analytical continuation in order to
derive the dynamic Josephson equations which the relative phases $\varphi _1$ and
$\varphi _2$ obey. To obtain the dissipative terms, it is sufficient to expand the
response functions $\alpha _{ik}$ and $\beta _{ik}$ up to linear terms in
frequency $\omega _n$.
\par
So, we look for the following first coefficients in the low-frequency
decomposition
\begin{gather*}
\alpha _{ik}(\omega _n)=-\alpha _{ik}^{(0)}-\alpha _{ik}^{(1)}|\omega _n|+\cdots
\\
\beta _{ik}(\omega _n)=-\beta _{ik}^{(0)}+\beta _{ik}^{(1)}|\omega _n|+\cdots
\end{gather*}
Accordingly, the expressions for the response functions in the imaginary-time
representation read as
\begin{gather*}
\alpha _{ik}(\tau)=-\alpha _{ik}^{(0)}\delta (\tau )+\alpha
_{ik}^{(1)}\frac{1}{\pi}\Bigl(\frac{\pi T}{\sin (\pi T\tau )}\Bigr)^2 +\cdots
\\
\beta _{ik}(\tau )=-\beta _{ik}^{(0)}\delta (\tau )-\beta
_{ik}^{(1)}\frac{1}{\pi}\Bigl(\frac{\pi T}{\sin (\pi T\tau )}\Bigr)^2 +\cdots
\end{gather*}
\par
First of all, we should note that zero harmonic $\omega _n=0$ in the $\alpha
_{ik}$ response is unimportant if $i=k$, and thus we can deal with the difference
$\tilde{\alpha}_{ii}(\omega _n)=\alpha _{ii}(\omega _n)-\alpha _{ii}(\omega
_n=0)$. In fact, this corresponds to the substitution $\alpha
_{ii}(\tau)=\tilde{\alpha} _{ii}(\tau)+\alpha _{ii}(0)\delta (\tau )$ into
effective action (\ref{f18}). The second term $\alpha _{ii}(0)\delta (\tau )$
yields a physically unimportant time- and phase-independent contribution to the
action, meaning a shift of the ground-state energy of a junction. For $i\neq k$,
this does not hold for. As we will see below, $\alpha _{i\neq k}^{(0)}$ and $\beta
_{ik}^{(0)}$ are connected with the Josephson potential energy and $\alpha
_{ik}^{(1)}$, $\beta _{ik}^{(1)}$ determine the dissipative properties of the
junction.
\par
Let us start from $\beta _{ik}^{(0)}\!$. For the sake of brevity, we present here
the expressions for $\alpha _{ik}$ and $\beta _{ik}$ in the case of a symmetric
junction with the identical mixtures on the both left-hand and right-hand sides of
the junction when $\Delta _{ik,\, l}=\Delta _{ik,\, r}=\Delta _{ik}$ and $n_{i\,
l}=n_{i\, r}=n_i$. The general case $l\neq r$ will be given in the Appendix. The
simple calculation yields
\begin{gather}
\beta _{11}^{(0)}=-2 n_1 f_{11}^+(\omega _n=0)= \nonumber
\\
\frac{n_1}{\pi}\,\frac{m_1^2\, (\Delta _{11}+m_2c_1c_2)}{(m_1\Delta
_{11}+2m_1m_2c_1c_2+m_2\Delta _{22})^{1/2}}\, , \nonumber
\\
\beta _{12}^{(0)}=\alpha _{12}^{(0)}=-2\sqrt{n_1n_2}\, f_{12}^+(\omega =0)=
\nonumber
\\
\frac{\sqrt{n_1n_2}}{\pi}\,\frac{m_1m_2\Delta _{12}}{(m_1\Delta
_{11}+2m_1m_2c_1c_2+m_2\Delta _{22})^{1/2}}\, ,  \label{f22}
\end{gather}
and the other quantities can be obtained with $1\leftrightarrows 2$.
\par
The calculation of $\beta _{ik}^{(1)}$ and $\alpha _{ik}^{(1)}$ is more
complicated:
\begin{gather}
\alpha _{11}^{(1)}=\beta _{11}^{(1)}=2n_1\,\frac{m_1}{4\pi
c_1c_2}\,\frac{c_1c_2+\Delta _{22}/m_2}{c_1+c_2}\, , \nonumber
\\
\alpha _{12}^{(1)}=\beta _{12}^{(1)}=2\sqrt{n_1n_2}\,\frac{1}{4\pi
c_1c_2}\,\frac{\Delta _{12}}{c_1+c_2}\, .\label{f23}
\end{gather}
Again all the remaining quantities are given by $1\leftrightarrows 2$. Note only
that all $\beta _{ik}^{(0)}$ and $\alpha _{i\neq k}^{(0)}$ remain finite and
nonsingular at the demixing point $\Delta _{11}\Delta _{22}=\Delta _{12}^2$ or
when one of the sound velocities vanishes $c_2=0$. On the contrary, both $\alpha
_{ik}^{(1)}$ and $\beta _{ik}^{(1)}$ diverge with approaching at demixing point as
$c_2\rightarrow 0$.

\section{JOSEPHSON EQUATIONS}
\par
To obtain the dynamics of the relative phases $\varphi _1$ and $\varphi _2$ in
real time, we now follow the standard procedure of analytical continuation.
Accordingly, the substitution of Matsubara frequencies $|\omega _n|\rightarrow
-i\omega$ in the Fourier transform of the Euler-Lagrange equations $\partial
S_{eff}/\partial\varphi _i(\tau)=0$,
\begin{gather*}
-\sum _{k=1,2}\widehat{U}_{ik}^{-1}\ddot{\varphi}_{k}(\tau) +2I_in_i\sin\varphi
_i(\tau )
\\
+ 2I_i\sum _{k=1,2}I_k\int d\tau '\,\Bigl[\alpha _{ik}(\tau -\tau
')\sin\Bigl(\varphi _i(\tau)-\varphi _k(\tau ')\Bigr)
\\
+ \beta _{ik}(\tau -\tau ')\sin\Bigl(\varphi _i(\tau)+\varphi _k(\tau
')\Bigr)\Bigr] =0 ,
\end{gather*}
will yield the real-time equation for the phase $\varphi _i$ dynamics.
\par
As we have found in the limit of the slowly varying phases $\varphi _{1,\, 2}$,
the response functions in the real-time representations go over to the following
decomposition
\begin{gather*}
\alpha _{ik}(t)=-\alpha _{ik}^{(0)}\delta (t)+\alpha _{ik}^{(1)}\delta '(t)+\cdots
\\
\beta _{ik}(t)=-\beta _{ik}^{(0)}\delta (t)-\beta _{ik}^{(1)}\delta '(t) +\cdots
\end{gather*}
Then, we can derive a couple of the Josephson equations which every relative phase
$\varphi _i(t)$ obeys in the real time $t$:
\begin{gather*}
\sum _{k=1,2}\widehat{U}_{ik}^{-1}\ddot{\varphi}_{k}(t) +2I_in_i\sin\varphi _i(t)
+\sum _{k=1,2} 2I_iI_k\times
\\
\Bigl[\alpha _{ik}^{(0)}\sin\Bigr(\varphi _i(t)-\varphi _k(t)\Bigl) +\beta
_{ik}^{(0)}\sin\Bigr(\varphi _i(t)+\varphi _k(t)\Bigl)\Bigr]
\\
+\sum _{k=1,2} 2I_iI_k\dot{\varphi} _k (t)\times
\\
\Bigl[\alpha _{ik}^{(1)}\cos\Bigl(\varphi _i(t)-\varphi _k(t)\Bigr) +\beta
_{ik}^{(1)}\cos \Bigl(\varphi _i(t)+\varphi _k(t)\Bigr)\Bigr] =0 .
\end{gather*}
Using relations from (\ref{f22}) and (\ref{f23}), we finally arrive at the desired
Josephson equations for two phases $\varphi _1(t)$ and $\varphi _2(t)$:
\begin{gather*}
\sum _{k=1,2}\widehat{U}_{ik}^{-1}\ddot{\varphi}_k +\sum _{k=1,2}4I_iI_k\alpha
_{ik}^{(1)}\dot{\varphi}_k\cos\varphi _i\cos\varphi _k
\\
+2I_in_i\sin\varphi _i +\!\!\sum _{k=1,2}\! 4I_iI_k\beta ^{(0)}_{ik}\sin\varphi
_i\cos\varphi _k =0 , \;\; i=1, 2.
\end{gather*}
\par
We can interpret these Josephson equations using a mechanical analogy. First, we
introduce the potential energy $U(\varphi _1,\,\varphi _2)$ of a junction
according to
\begin{equation*}
U(\varphi _1,\,\varphi _2)=-\!\!\sum\limits _{i=1,\, 2}E_i\cos\varphi _i
+\frac{1}{2}\sum\limits _{i,k=1,2}\!\varepsilon _{ik}\cos\varphi _i\cos\varphi _k
\end{equation*}
with the energy coefficients determined by
\begin{gather*}
E_i=2I_in_i \;\;\text{and}\;\; \varepsilon _{ik}=4I_iI_k\beta _{ik}^{(0)} ,\;\;
(i,\, k=1,\, 2) .
\end{gather*}
Then we can introduce the Lagrangian $L=K-U$ as a difference between the kinetic
$K$ and potential $U$ energies
\begin{equation*}
L=K-U=\sum\limits _{i, k= 1, 2}\frac{\widehat{U}_{ik}^{-1}}{2}
\dot{\varphi}_i\dot{\varphi}_k -U(\varphi _1,\,\varphi _2) .
\end{equation*}
The dissipation energy effect can be described by introducing the dissipative
Rayleigh function according to
\begin{gather*}
R(\dot{\varphi}_1,\,\dot{\varphi}_2)=\frac{1}{2}\sum\limits _{i,k=1,2}
r_{ik}(\varphi _1,\,\varphi _2)\,\dot{\varphi}_i\dot{\varphi}_k
\\
=\frac{1}{2}\sum\limits _{i,k=1,2}r_{ik}\cos\varphi _i\cos\varphi
_k\,\dot{\varphi}_i\dot{\varphi}_k ,\;\text{and}\;\; r_{ik}=4I_iI_k\alpha
_{ik}^{(1)}.
\end{gather*}
Finally, the Josephson equations can be written in the general form as
\begin{equation*}
\frac{d}{dt}\,\frac{\partial L}{\partial\dot{\varphi} _i}-\,\frac{\partial
L}{\partial\varphi _i}=-\,\frac{\partial R}{\partial\dot{\varphi}_i}\, , \;\;\;
i=1, 2,
\end{equation*}
where in accordance with Eq.~(\ref{f16}),
\begin{equation*}
\dot{\varphi}_i(t)=-\delta\mu _i(t)=\mu _{i,\, l}-\mu _{i,\, r} .
\end{equation*}
\par
The dissipative function $R$ has a sense of the energy dissipation power in the
system. This is obvious from the following equation,
\begin{equation*}
\frac{dH}{dt}=\frac{d}{dt}\Bigl(\sum\limits _i\dot{\varphi}_i\frac{\partial
L}{\partial\dot{\varphi}_i}-L\Bigr)=-\sum _i\dot{\varphi}_i\frac{\partial
R}{\partial\dot{\varphi}_i}=-2R ,
\end{equation*}
which means that the energy dissipation power equals the double dissipative
function. Since the energy dissipation must result in decreasing the total energy
$H=K+U$ of the system, the dissipative function $R$ must be a positively
determined matrix $R>0$, i.e.,
\begin{equation*}
r_{11}>0 \;\;\;\text{and}\;\;\; r_{11}r_{22}>r_{12}r_{21} .
\end{equation*}
We are persuaded that this is true from
\begin{equation*}
\frac{r_{11}r_{22}}{r_{12}r_{21}}-1=\frac{\alpha _{11}^{(1)}\alpha
_{22}^{(1)}}{\alpha _{12}^{(1)}\alpha
_{21}^{(1)}}-1=\frac{m_1m_2c_1c_2(c_1+c_2)^2}{\Delta _{12}^2}>0 .
\end{equation*}
Note that the condition $R>0$ gets broken simultaneously with the condition
$c_{1,2}>0$ necessary for the stability of a mixture against its demixing. In
addition, we also disclose a symmetry of kinetic dissipative coefficients
$r_{ik}=r_{ki}$ in accordance with the Onsager principle.
\par
It is interesting that the point of demixing instability $c_1c_2=0$ is not
singular for the potential energy coefficients $\varepsilon _{ik}$. On the
contrary, the dissipative coefficients $r_{ik}$ become infinite. The latter means
that the Josephson dynamics should slow down and demonstrate an enhancement of
decoherence and damping of the Josephson oscillations in the vicinity of the phase
demixing. From the mechanical point of view the low-frequency dynamics of a
Josephson junction in a Bose-condensed mixture can be described as a system of two
coupled particles or pendula moving or oscillating in a viscous medium in a
periodic potential relief.
\par
In the lack of hybridization between the different atom species, i.e., when
$\Delta _{12}=0$, the crossed terms in the response functions $\alpha _{i\neq k}$
and $\beta _{i\neq k}$ vanish as well. As both $\alpha _{i\neq k}=0$ and $\beta
_{i\neq k}=0$, the Josephson equations split into two decoupled equations for each
component of a mixture. In this case no interference in the mass currents appears
and the response functions together with nonzero diagonal coefficients $\alpha
_{ii}$, $\beta _{ii}$ go over to the quantities corresponding to the case of a
single-component Bose-condensed gas \cite{Mei,Ba}.
\section{CONCLUSION}
\par
To summarize, in this paper we have used a functional integration method for the
model of a tunneling Hamiltonian in order to analyze the energy dissipation
effects in the dynamics of a pointlike Josephson junction between two weakly
nonideal Bose-condensed gas mixtures in the thermal equilibrium. The transmission
of particles of each component of a mixture across the junction is described by
two different tunneling amplitudes. The effective action and response functions
that describe the dynamics of two relative phases $\varphi _1$ and $\varphi _2$
corresponding to each condensed component of a mixture are found. The
quasiclassical Josephson equations for the relative phases are derived from the
low-frequency decomposition of the response functions.
\par
The dynamics of a pointlike junction in a mixture displays a dissipative Ohmic
nature. The energy dissipation effects result from the noncondensate excitations
and appear in second order in the tunneling amplitudes. The latter fact favors low
damping rates of the Josephson oscillations in the pointlike junctions. A growth
of the temperature leads to decreasing the Josephson energy and to increasing the
energy dissipation power. The closeness to the phase separation of a mixture
enhances the Ohmic character of the phase dynamics. The dissipative Rayleigh
function is determined.
\par
On the whole, the Josephson phase dynamics in binary mixtures is described by two
coupled equations. This means, in particular, an existence of two Josephson
frequencies for small oscillations of the phases around $\varphi _1=0$ and
$\varphi _2=0$. Since $r_{ik}\neq 0$, the oscillations are weakly damped.
Emphasize that the interference between the Josephson and dissipative Ohmic
components of a mass current for each type of particles starts only from the
second-order terms in the tunneling amplitudes and, eventually, due to the
presence of the noncondensate fractions. The interference entails, in particular,
that the maximum amplitude of the Josephson current of one species atoms depends
on the relative phase difference of the second component of a mixture. In
addition, it becomes possible that the imbalance in the chemical potential of one
component of a mixture can induce also the Ohmic contribution into the mass
current of the other component. We believe these aspects deserve a further study.

\section*{ACKNOWLEDGMENTS}
\par
This study is supported in part by the RFBR grant No. 10-02.00047a.
\section*{APPENDIX}
\par
Here we present the general asymmetric case of a junction if $n_{i\, l}\neq n_{i\,
r}$ and $ \Delta _{ik,\, l}\neq\Delta _{ik,\, r}$. For the Josephson potential
energy
\begin{equation*}
U(\varphi _1,\,\varphi _2)=-\!\sum\limits _{i=1,\, 2}E_i\cos\varphi _i
+\frac{1}{2}\sum\limits _{i,\, k=1,\, 2}\!\varepsilon _{ik}\cos\varphi
_i\cos\varphi _k ,
\end{equation*}
we find
\begin{equation*}
E_i=2I_i\sqrt{n_{i\, l}n_{i\, r}}\;\;\; (i=1,\, 2)
\end{equation*}
and the next terms
\begin{gather*}
\varepsilon _{11}=\frac{2I_1^2m_1^2}{\pi}\Bigl[\frac{n_{1l}(\Delta _{11,r}
+m_2c_{1r}c_{2r})}{(m_1\Delta _{11,r} +2m_1m_2c_{1r}c_{2r}+m_2\Delta
_{22,r})^{1/2}}
\\
+ \frac{n_{1r}(\Delta _{11,l}+m_2c_{1l}c_{2l})}{(m_1\Delta
_{11,l}+2m_1m_2c_{1l}c_{2l}+m_2\Delta _{22,l})^{1/2}}\Bigr] ,
\\
\varepsilon _{22}=\frac{2I_2^2m_2^2}{\pi}\Bigl[\frac{n_{2l}(\Delta
_{22,r}+m_1c_{1r}c_{2r})}{(m_1\Delta _{11,r}+2m_1m_2c_{1r}c_{2r}+m_2\Delta
_{22,r})^{1/2}}
\\
+\frac{n_{2r}(\Delta _{22,l}+m_1c_{1l}c_{2l})}{(m_1\Delta
_{11,l}+2m_1m_2c_{1l}c_{2l}+m_2\Delta _{22,l})^{1/2}}\Bigr] .
\end{gather*}
The other two nondiagonal coefficients $\varepsilon _{12}=\varepsilon _{21}$ are
given by the expression
\begin{gather*}
\varepsilon _{12}=
\frac{2I_1I_2}{\pi}\Bigl[\frac{\sqrt{n_{1l}n_{2l}}}{c_{1r}c_{2r}}\frac{\Delta
_{12,r}}{c_{1r}+c_{2r}}+\frac{\sqrt{n_{1r}n_{2r}}}{ c_{1l}c_{2l}}\frac{\Delta
_{12,l}}{c_{1l}+c_{2l}}\Bigr] .
\end{gather*}
\par
For the kinetic coefficients in the dissipative Rayleigh function
\begin{equation*}
R(\dot{\varphi}_1,\,\dot{\varphi}_2)=\frac{1}{2}\sum\limits _{i=1,\,
2}r_{ik}\cos\varphi _i\cos\varphi _k\,\dot{\varphi}_1\dot{\varphi}_2\, ,
\end{equation*}
we have
\begin{gather*}
r_{11}=\frac{I_1^2m_1}{\pi}\Bigl[\frac{n_{1l}}{c_{1r}c_{2r}}\,\frac{c_{1r}c_{2r}
+\Delta_{22,r}/m_2}{c_{1r}+c_{2r}}
\\
+\frac{n_{1r}}{c_{1l}c_{2l}}\,\frac{c_{1l}c_{2l}
+\Delta_{22,l}/m_2}{c_{1l}+c_{2l}}\Bigr],
\\
r_{22}=\frac{I_2^2m_2}{\pi}\Bigl[\frac{n_{2l}}{c_{1r}c_{2r}}\,\frac{c_{1r}c_{2r}
+\Delta_{11,r}/m_1}{c_{1r}+c_{2r}}
\\
+\frac{n_{2r}}{c_{1l}c_{2l}}\,\frac{c_{1l}c_{2l}
+\Delta_{11,l}/m_1}{c_{1l}+c_{2l}}\Bigr].
\end{gather*}
The other two nondiagonal coefficients $r _{12}=r _{21}$ can be found from the
expression
\begin{gather*}
r_{12}=\frac{I_1I_2}{\pi}\Bigl[\frac{n_{1l}n_{2l}}{c_{1r}c_{2r}}\,\frac{\Delta
_{12,r}}{c_{1r}+c_{2r}}+\frac{n_{1r}n_{2r}}{c_{1l}c_{2l}}\,\frac{\Delta
_{12,l}}{c_{1l}+c_{2l}}\Bigr] .
\end{gather*}
\par
The expressions derived above for the dynamical coefficients in the Josephson
equations governing the phase differences $\varphi _1(t)$ and $\varphi _2(t)$
across a pointlike junction allow us to describe the low-frequency dynamics in the
asymmetric case of Bose-condensed gas mixtures with the different densities of the
atom species and with the different order parameters. The stability of mixtures
against phase separation and the positive definiteness of the dissipative Rayleigh
function imply the fulfillment of conditions $c_{1l}$, $c_{2l}>0$ and $c_{1r}$,
$c_{2r}>0$.

\end{document}